# A theoretical view on unimolecular rectification

**R Stadler, V Geskin, and J Cornil**
Laboratory for Chemistry of Novel Materials, University of Mons-Hainaut, Place du Parc 20, B-7000 Mons, Belgium

E-mail: r.stadler@averell.umh.ac.be

**Abstract:** The concept of single molecule rectifiers proposed in a theoretical work by Aviram and Ratner in 1974 was the starting point of the now vibrant field of molecular electronics. In the meantime, a built-in asymmetry in the conductance of molecular junctions has been reported at the experimental level. In this contribution, we present a theoretical comparison of three different types of unimolecular rectifiers: i) systems where the donor- and acceptor-part of the molecules are taken from charge-transfer salt components; ii) zwitterionic systems; and iii) Tour wires with nitro substituents. We conduct an analysis of the rectification mechanism in these three different types of asymmetric molecules on the basis of parameterized quantum-chemical models as well as with a full non-equilibrium Green's function / density functional theory (NEGF-DFT) treatment of the current/voltage characteristics of the respective metal/molecule/metal junctions. We put a particular emphasis on the prediction of rectification ratios (RR), which are crucial for the assessment of the technological usefulness of single molecule junctions as diodes. We also compare our results with values reported in the literature for other types of molecular rectification, where the essential asymmetry is not induced by the structure of the molecule alone but either by a difference in the electronic coupling of the molecule to the two electrodes or by attaching alkyl chains of different lengths to the central molecular moiety.

**1. Introduction : Molecular electronics and the original proposal of Aviram and Ratner**

In 1974, Aviram and Ratner introduced a concept for a rectifier based on the use of a single organic molecule [1]. This diode consisted of a molecule featuring a donor and acceptor group, which are separated by a sigma-bonded tunneling bridge (see figure 1). In their theoretical work, the transfer of the electron from the cathode to the anode via the molecule was divided into three steps, namely the hopping of the electron from the cathode to the acceptor part of the molecule, then from the acceptor to the donor moieties inside the molecule, and finally from the donor part of the molecule to the anode. In this scenario, the emphasis is put on resonances between energy levels in the system rather than on the electron transfer rate inside the molecule. Namely, it is assumed that the energetic alignment of the Fermi level of the electrodes with respect to the localized frontier molecular orbitals on the donor and acceptor parts is responsible for the rectification effect.

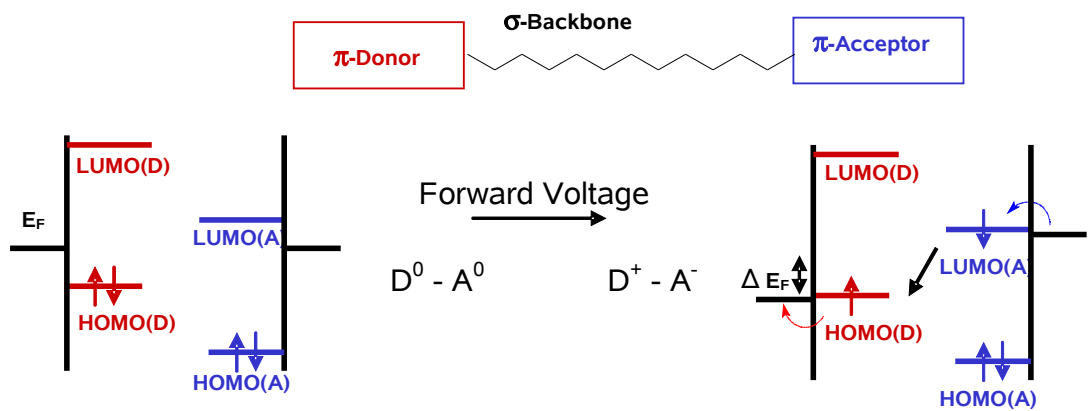

**Figure 1.** Illustration of the original Aviram-Ratner concept in Ref 1. The key feature is the energetic alignment of the Fermi level $E_F$ of the electrodes with respect to the localized HOMO and LUMO levels on the π-donor and acceptor units, respectively. In this simple model, the actual energetic alignment at zero bias determines which voltage direction gives rise to the largest current.

Thirty years later, molecular electronics has become a very active field of research [2-5]; the main incentive for this field can be found in economic necessity, since despite of continuous achievements in miniaturization of CMOS technology, fundamental difficulties will be faced when approaching the nanoscale. Some difficulties, such as the irreproducibility of detailed atomic structures based on silicon, can be solved by replacing some components of electric circuits by organic molecules, which have well-defined structures that can be mass produced by means of chemical synthesis. The rich variety of structures and functionalities in organic chemistry offers a huge toolbox for the implementation of logic functions or storage elements at the nanoscale.

Interest for electron transport through a single-molecule inserted between two nanoscale contacts - which is the most important physical process in molecular electronics – has also intensified within the last ten years due to : (i) recent progress in the experimental techniques for manipulating and contacting individual molecules [6-9]; and (ii) the availability of first-principles methods to describe the electrical properties of single molecule junctions at the atomic scale with high accuracy [9-15]. These theoretical methods are usually based on density functional theory (DFT) in combination with a non-equilibrium Green's function formalism [16] implemented in different ways.

With these new experimental and theoretical tools at hand, attempts to measure [17-21] and compute [22-24] the characteristic properties of single molecule rectifiers have recently intensified, not only for historical reasons but also because diodes are the simplest possible electronic devices that can be used for the implementation of logic circuits and memory elements [5].

In this contribution, we provide a theoretical framework for the characterization of different classes of single molecule rectifiers. We choose as target systems for our comparison : i) systems with separated donor (D) and acceptor (A) parts, where the two moieties of the molecules are taken from charge-transfer salt components, ii) zwitterionic systems; and iii) Tour wires with nitro substituents (all structures are shown in figure 2). We have performed quantum-chemical calculations at the Austin model (AM1) level [25] to study the rearrangement of the electronic structure of isolated molecules in response to an electric field, and density functional theory calculations coupled to a non-equilibrium Green's function technique (NEGF-DFT) [12,16,26] for an explicit description of the bias-dependent electron transport through metal-molecule-metal junctions. We address different approaches for defining and calculating the rectification ratio, which is a key quantity for an experimental assessment of the quality of single molecule rectifiers; we also discuss the origin of different shapes in the current/voltage curves in relation to the nature of the molecules and the way they are connected to the electrodes.

Our article is entirely focused on the electron transport regime of coherent tunneling as is most of the recent theoretical literature on single molecule rectification, at least when NEGF-DFT calculations are employed (see e. g. [23],[24]). The argument for this is usually that neither electrons nor nuclei have enough time to relax if the molecules are strongly coupled to metal electrodes by thiol anchor groups. It is debatable whether this contradicts the original Aviram Ratner model [1], where relaxations have been included as corrections but were not crucial to the proposed mechanism. In Ref. [1] the coupling between the electrodes and the molecule was only treated in a very approximative way, which did not allow to distinguish quantitatively between different transport regimes.

## 2. Computational approaches

The two methods that we use in this work, namely AM1 for the isolated molecules and NEGF-DFT for the electrode/molecule/electrode systems vary significantly in three aspects. First, AM1 is a wave-function approach that allows for the description of the ground and excited electronic states in the framework of a configuration interaction (CI) scheme. In contrast, with NEGF-DFT, the ground-state electron density under the influence of a given external potential (when including an applied electric field) is the central variable. Second, AM1 is a parameterized approach, which makes it computationally very efficient for systems for which reliable atomic parameters are available, *i.e.*, for the atoms typically found in organic molecules but only for a rather small selection of metallic atoms. In contrast, DFT calculations do not require parameters fitted to experiments, [27] which has the advantage that the issue of transferability of such parameters to different systems or different physical boundary conditions never arises; however, this comes at the cost of very demanding computational times. The third difference is the most crucial one for this work: with AM1, we shift charges between the donor and acceptor sides inside the isolated molecules, whereas with NEGF-DFT, we solve the non-equilibrium problem of a current passing from one electrode to another through the molecule.

In order to compute the current with the NEGF-DFT formalism, the transmission function T has to be integrated over the bias window for a junction with its electron density polarized by the generated electric field [12,26]. For the single molecular junctions considered here, T was calculated using a general non-equilibrium Green's function formalism for phase-coherent electron transport, [16] where both the Green's function of the scattering region and the self-energies describing the coupling to the semi-infinite electrodes were evaluated in terms of localized basis functions [27]. The MO evolution as a function of the bias, as computed with NEGF-DFT, is obtained by projecting the eigenstates of the semi-infinite junction on the sub-Hamiltonian defined by the part of the basis set that is localized exclusively on the molecule [12,23]. We stress that the MO eigenenergies calculated in this way do not coincide in general with the peaks in the transmission functions when the molecules are rather strongly coupled to the electrodes, as it is the case for all systems in the present study. Energetic shifts are observed as a result of the hybridization between the MOs and the metallic surface states; both the magnitude of these shifts and the width of the transmission peaks are proportional to the amount of coupling between the molecular levels and the electrodes. In our calculations, the supercell for the scattering region is defined by 3x3 atoms in the direction perpendicular to the transport direction and contains three surface layers on each side of the molecule. We found that a 3x3 k-point grid is needed for the sampling in the transverse Brillouin plane in order to obtain sufficiently accurate results for T and the current intensity. The atomic configurations used for the molecular structures inserted in the junctions are those relaxed at zero bias at the AM1 level and attached to the gold electrodes with thiolate groups; the sulfur atoms have been placed in a hollow position with respect to the surface plane at a vertical distance of 1.7 Å [28,29].

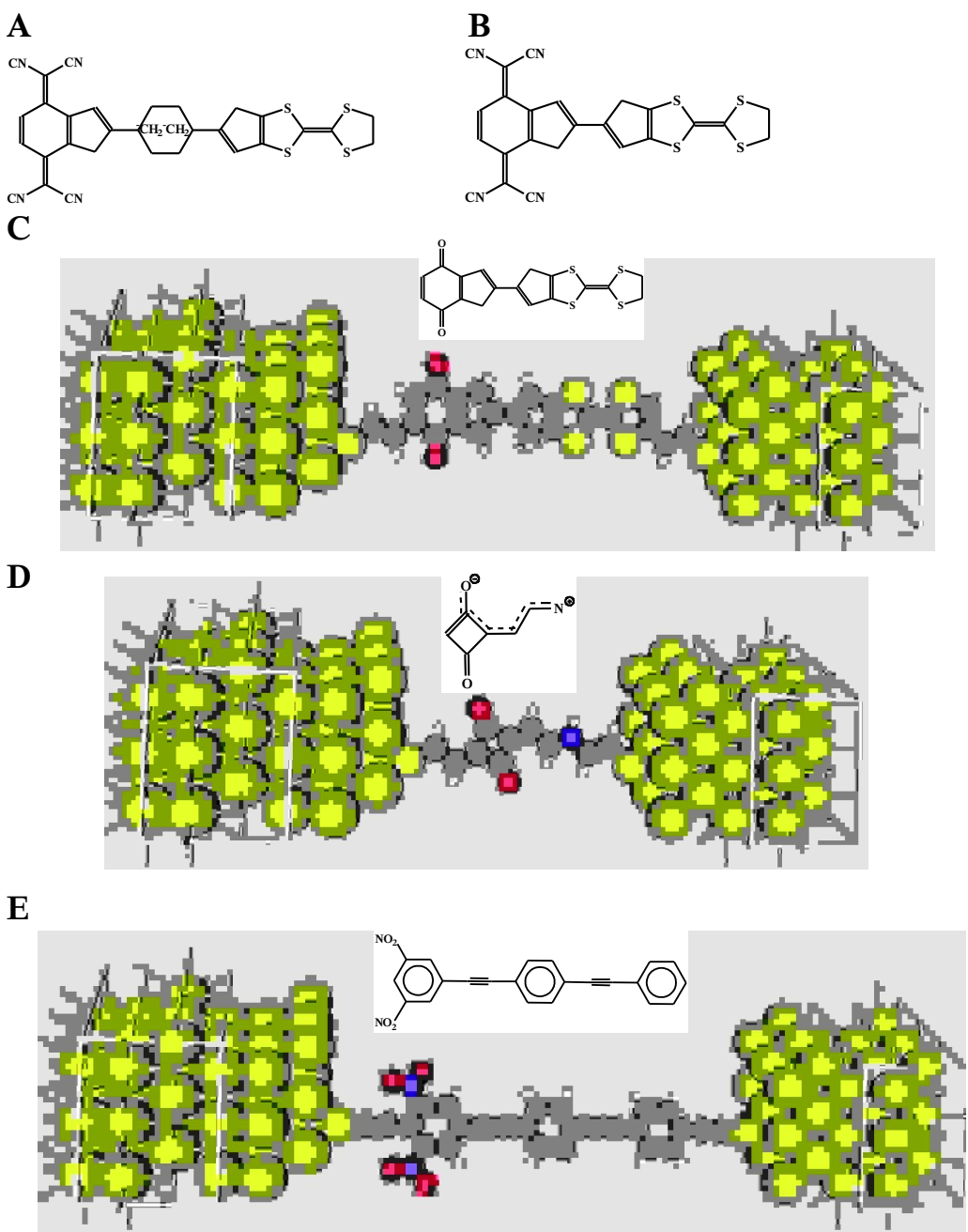

**Figure 2.** An overview of the systems under study. For molecules A and B, only AM1 calculations for intra-molecular electron transfer have been performed. The systems C, D, and E

have been studied with both the AM1 and NEGF-DFT methods. For geometrical reasons, spacers made of two carbon atoms have been put on each side between the central molecule and the thiol anchoring groups that establish the electronic connection with the gold electrodes. In order to preserve the conjugation patterns of the respective molecules, these two atoms are linked by a double bond for structures C and D and by a triple bond for E.

In figure 2, we provide an outlook of the molecular structures investigated in this contribution. Molecule A is the original single molecule rectifier suggested by Aviram and Ratner [1]. The reasoning in choosing this particular molecule was to combine a very strong donor with a very strong acceptor via an insulating and rigid aliphatic bridge. Tetracyanoquinodimethane (TCNQ) and tetrathiafulvene (TTF), the acceptor and donor moieties in this molecule, respectively, are known in the field of charge-transfer complexes [30] to form one of the strongest donor-acceptor pair. Our results show (see next section) that the removal of the bicyclooctane bridge, that is moving from molecule A to B, does not significantly alter the intra-molecular charge transfer characteristics. Another derivatization can be achieved by replacing TCNQ with the weaker acceptor para-benzoquinone (molecule B → C). This is expected to change the electron transfer in a quantitative manner but to keep unaffected the basic mechanism of the process. In order to keep the computational efforts of this study within reasonable limits, we performed NEGF-DFT calculations for this class of compounds only for structure C while the intra-molecular charge transfer has been characterized for all three molecules with AM1.

Amino-vinyl squaraine (molecule D) is the monomer of a semiconducting polymer with a very small band gap [31] and was chosen in view of its zwitterionic structure. In recent conductance measurements on rather large molecules, quite high rectification ratios (RR) were obtained and attributed to the physical properties of the zwitterion, thus motivating the focus on this compound. Molecule E is a so-called Tour wire [7] with strategically placed electron-withdrawing nitro substituents. In a recent theoretical study, the influence of the position of the nitro group on the performance of π-conjugated unimolecular rectifiers was analyzed and explained [32]. We have performed NEGF-DFT calculations for both molecules D and E.

**3. Spatial asymmetry and molecular properties influencing rectification**

*3.1. A general scheme for the classification of unimolecular rectifiers*

In order for a single molecule junction to exhibit rectifying characteristics in its current dependence versus the polarity of the applied bias, spatial asymmetry in some part of the atomic structure of the junction is an obvious key requirement. There are, however, three ways allowing for such an asymmetry to be introduced [21]: i) the coupling at the interface between the molecule and the two electrodes might differ in the sense that the molecule is e.g. chemisorbed on one side and physisorbed on the other side [33,34]; ii) a central electroactive moiety might be placed asymmetrically inside the junction, e.g. by being connected with alkyl chains of different lengths [35,36]; and iii) the electron density of the relevant molecular orbitals (MOs) in the conjugated core of the molecule is polarized by donor and acceptor functionalities. We focus here mostly on case iii) which is usually referred to as "truly" unimolecular rectification. The classification scheme that we introduce in this section is, however, general.

In figure 3, we show schematically the MO level crossings which have to occur to yield an intra-molecular electron jump in an asymmetric molecule when an external electric field is applied; the latter can be modeled e.g. by placing two point charges Q of opposite sign (so-called sparkles) at a rather large distance from the molecule. In a recent study, we have demonstrated that the rectification ratios (RR) derived from such intra-molecular jumps can give a good indication of the rectifying properties of single molecule junctions [32]; the analysis performed hereafter will build up on this previous work. Such diagrams can also be derived from NEGF-DFT calculations by projecting the eigenvalues of the self-consistent Hamiltonian of the whole junction over its molecular part [23].

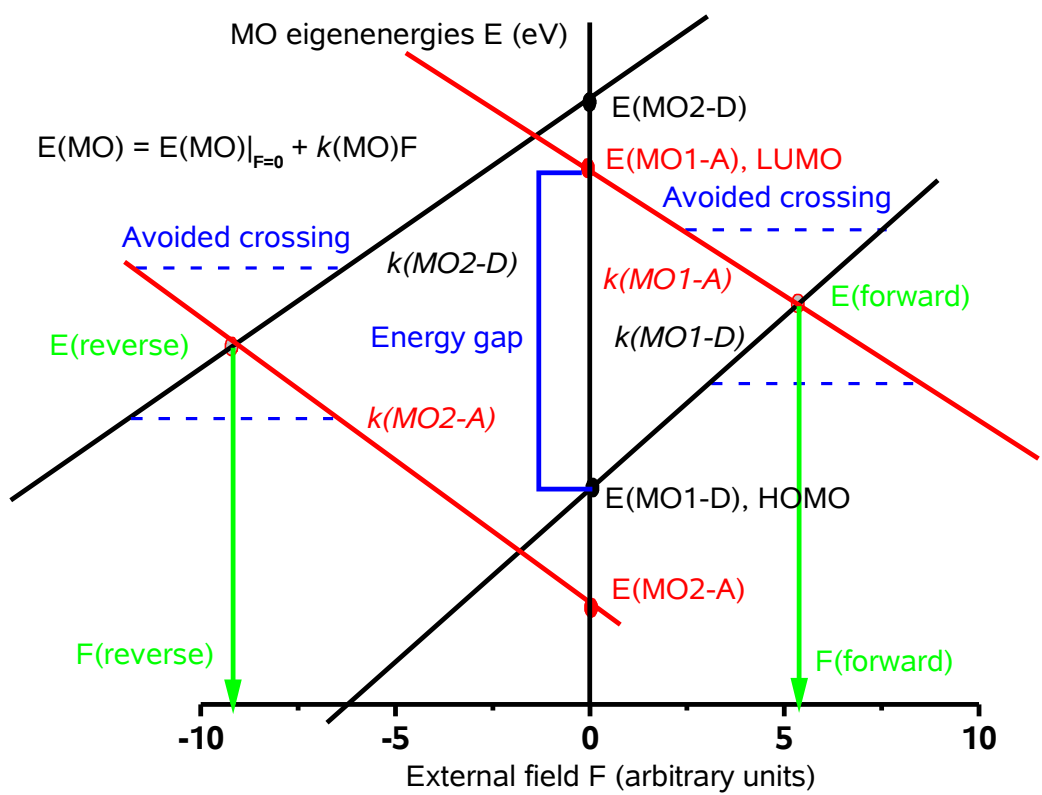

**Figure 3.** Sketch of the parameters characterizing intra-molecular electron jumps in asymmetric molecules. For each frontier orbital, its eigenenergy at zero bias $E(MO)|_{F=0}$ and the slope $k$ of its dependence versus the external electric field F has to be considered. Avoided crossings occur at the points E(forward) and E(reverse) that can be used to estimate RR (=F(forward)/F(reverse)) or (=Q(forward)/Q(reverse)).

Assuming that there are four frontier orbitals involved in the electron jumps in both directions, there are eight parameters which determine the values for the threshold voltages F(forward) and F(reverse), whose ratio can be taken as an indicator for RR (see figure 3): the four energies E(MO1-A), E(MO1-D), E(MO2-A) and E(MO2-D) of the MOs at zero bias, and their evolution with the external electric field that defines the slopes $k$(MO1-A), $k$(MO1-D), $k$(MO2-A) and $k$(MO2-D), respectively. In this diagram, MO1-A and MO2-A are localized on the acceptor side of the molecule and MO1-D and MO2-D on its donor part.

The slopes $k$(MO) depend on the spatial position and/or localization of the respective MOs, whose importance has been previously recognized in scanning tunnelling microscopy [37] (where the capacitive coupling of the central molecule to the tip and the surface is asymmetric) and for monolayers of molecules with alkyl groups of different lengths, both with a single [35] or two active MOs [38] in the central moiety. In a recent DFT-based study, these slopes have been rationalized in terms of the polarizability of the two halves of an asymmetric molecule [39] (including the gold clusters they were attached to and the resulting screening effects); such slopes were also exploited to explain the influence of the spatial position of nitro substituents on the properties of π-conjugated unimolecular rectifiers with NEGF-DFT calculations [32].

The eigenenergies of the localized MOs at zero bias are directly linked to the acceptor and donor strength of the two parts of the asymmetric molecule. They have been the key ingredient of the first Aviram Ratner proposal [1] and of a simplistic design scheme that has been derived from it [22]. The idea is that a strong acceptor and a strong donor would result in energetically low values for E(MO1-A) and E(MO2-A) and high values for E(MO1-D) and E(MO2-D), respectively that would maximize the ratio Q(reverse)/Q(forward) in figure 3. This argument motivated us to study structures A to C displayed in figure 2. Another consequence of a high E(MO1-D) and a low E(MO1-A) is a low energy gap, thus pointing to the interest of considering structure D. Finally, structure E was picked for our comparison since it is an example of the way RR can be optimized by a strategic placement of the nitro groups to control $k$(MO), as discussed in Ref. [32].

*3.2. Systems with strong donor and acceptor groups*

In figure 4a, we show the field evolution of the AM1 electronic structure of compounds A, B, and C, with the external electric field generated by point charges with opposite signs positioned 6 Å away from the terminal atoms of the molecules. The threshold values of the sparkle charges at which electron jumps occur have been marked by horizontal lines; these jumps are detected via a Mulliken population analysis at the AM1/CI level when the charge on the acceptor moiety becomes equal to -1|e|. The bicyclooctane bridge, which is the only structural feature that distinguishes molecule A from B, has no pronounced influence on either the zero bias energy gap or the threshold voltages. We attribute this to the preexisting weak conjugation in molecule B in the absence of the bridge, as supported by the localization of the frontier orbitals over the donor and acceptor parts. On the other hand, molecule C has a weaker acceptor group, which translates into an increased gap size and requires larger sparkle charges for the electron transfer to occur; we will show in the next section that this tends to reduce somewhat RR. All molecular orbitals participating in the electron jumps in both directions have the same spatial distribution pattern in the three molecules (see the insets of figure 4a), thus indicating that it is safe to limit our NEGF-DFT calculations to molecule C for understanding the rectifying mechanism for this class of systems.

The MO evolution as a function of the bias, as computed with NEGF-DFT, is depicted in figure 4b for molecule C. We stress that the scale of the voltage range in figure 4b is about one order of magnitude smaller than for the intra-molecular electron jumps in figure 4a, since we have considered the range of voltages which is usually applied in experimental measurements [18,19]. The two levels which play a major role in this bias regime are the HOMO and LUMO levels of the entire system which get closer energetically in the forward bias direction and move apart when the direction of the applied voltage is reversed. Note that the two levels have the same spatial distribution as those calculated at the AM1 level (see figure 4a). There are, however, additional states in figure 4b, associated to the ethylene spacers and the thiolate anchors connecting the molecule to the gold electrodes, which do not enter into the bias window. Note also that there is a maximum in the HOMO energy at V = 0V and that the HOMO follows the border of the bias window before entering into it. This behavior is due to the charging of the molecule which diminishes the effect of the external electric field. The charging is not only felt by the HOMO but also by all MOs localized on the same part of the molecule which therefore move parallel to the HOMO level in the energy diagram. We observe the same feature for molecule D in figure 5.

The transmission functions in Figure 4c show that the LUMO is rather weakly coupled to the electrodes, leading to peaks with low intensity for both zero and finite voltages. It is mostly the HOMO level which yields a pronounced peak close to the Fermi level and dominates the asymmetry in the I/V curves and RR, as discussed in the next chapter.

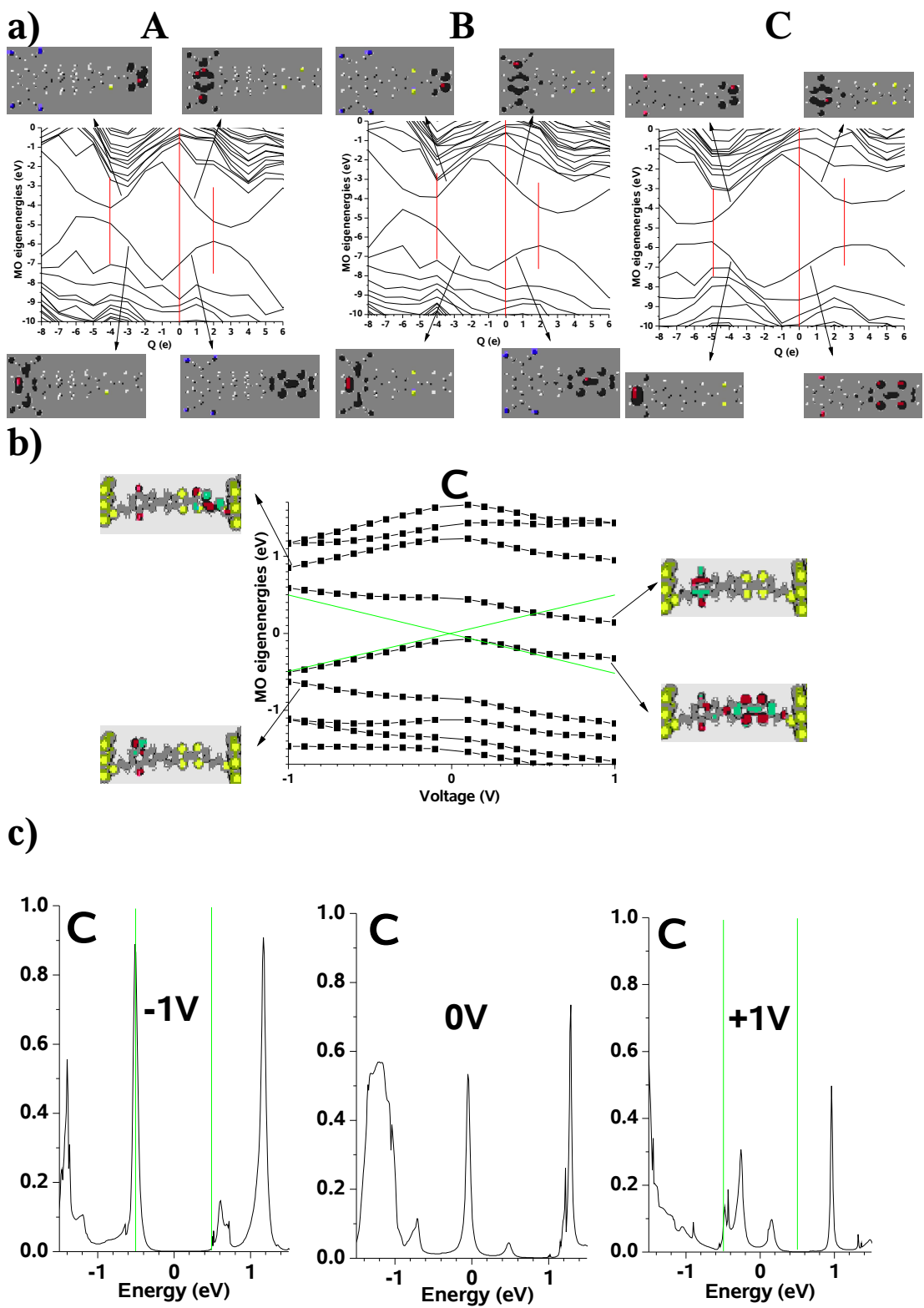

**Figure 4.** Evolution of the MO eigenenergies as a function of: a) the sparkle charges Q for structures A, B, and C at the AM1 level (the vertical lines indicate the threshold for a full electron

transfer); b) the external bias for structure C, as calculated with NEGF-DFT; c) Transmission functions for molecule C for voltages of -1V, 0V and +1V. The bias window is shown with bright/green lines in b) and c) and the frontier MOs as insets in a) and b). In the latter, the snapshot of the LCAO (Linear Combination of Atomic Orbitals) pattern corresponds to the charge/voltage indicated by the arrow.

*3.3. Zwitterionic structure*

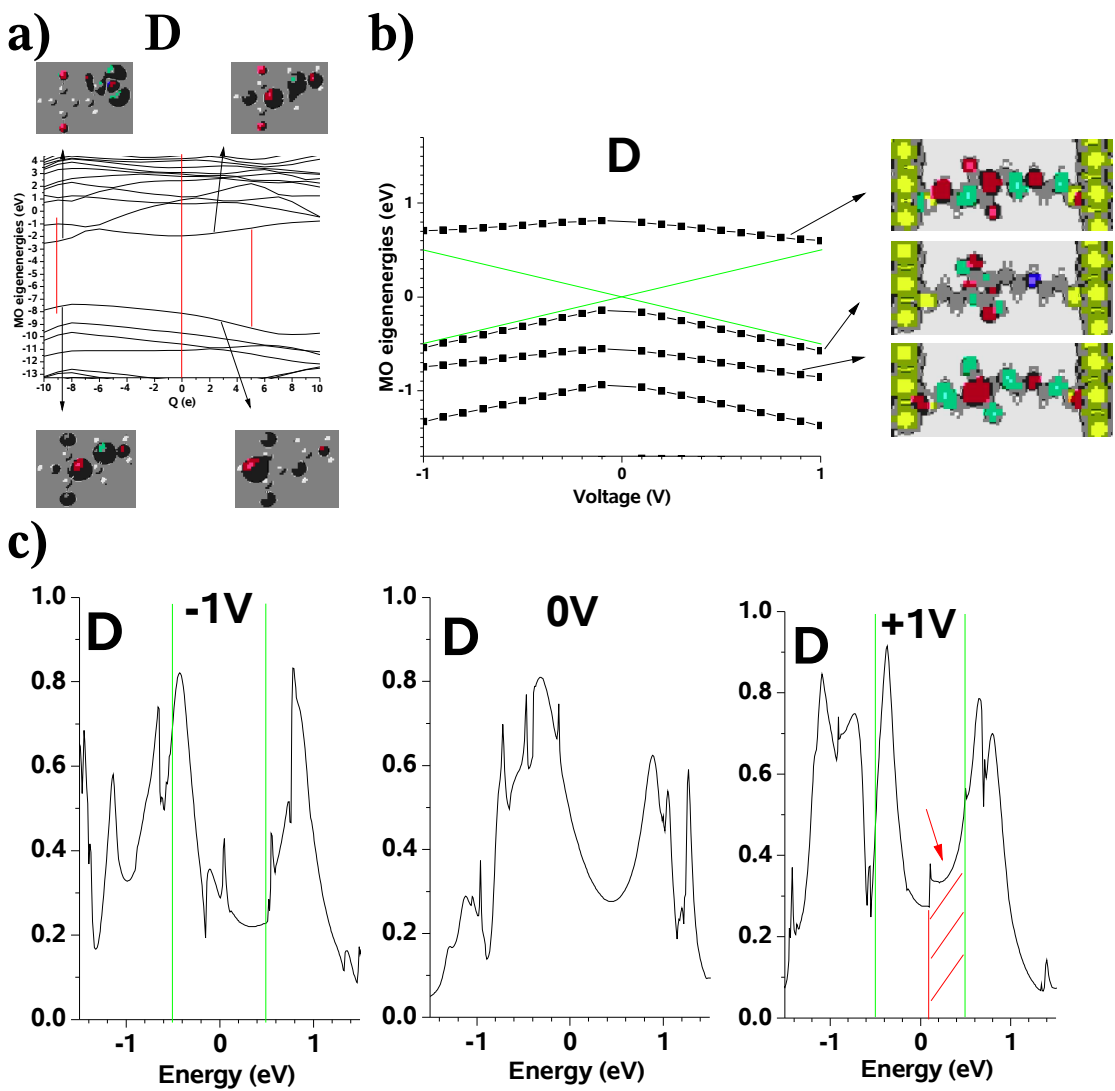

**Figure 5.** Evolution of the MO eigenenergies with the external field and transmission functions at different voltages for molecule D. See details in the caption of figure 4. The vertical lines in a) refer to Mulliken charges of 1|e| on the donor and acceptor sides; note, however, that there are no discrete jumps in this case.

Zwitterions such as molecule D are considered to be good candidates for single molecule rectifiers since charge separation has already occurred for the forward bias and would have to be reversed in the opposite direction [19,21]. This charge-separated state is, however, just one of a few possible mesomers in the chemical structure. The AM1 results reported in figure 5a show that there is no discrete electron jump between the two parts of the molecule when a forward bias is applied and that at the reverse bias charge transfer occurs at very high voltages when compared to the other molecules. Electron transfer is promoted by a continuous change in the localization pattern of the HOMO level induced by the polarization of the molecular orbitals and hence energetic shifts of the involved atomic orbitals. In figure 5b, only a rather weak asymmetry of the level shifts computed with NEGF-DFT can be detected with respect to the polarity of the applied voltage, except for the LUMO that comes slightly closer to the bias window in the forward direction. As a consequence, the transmission spectrum inside the bias window for +1V shows an enhanced shoulder (marked by an arrow and shaded area in figure 5c), which is lacking at -1V. The insets of figure 5b illustrate that the levels dominating the transmission functions (LUMO and HOMO-1) are delocalized over the whole molecule and are rather strongly coupled to both electrodes.

*3.4. Tour wires with nitro substituents*

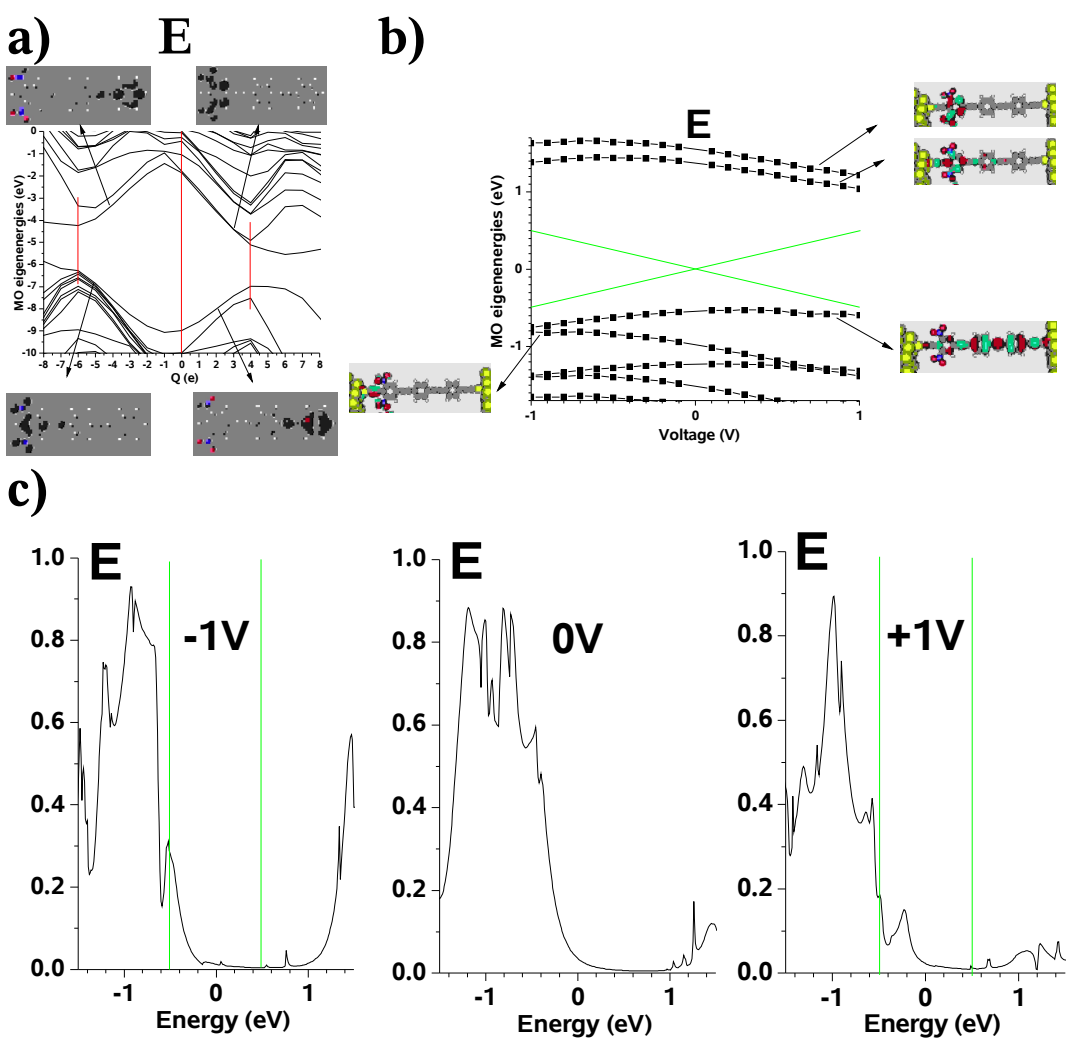

**Figure 6.** Evolution of the MO eigenenergies with the external field and transmission functions at different voltages for molecule E. See details in the caption of figure 4.

One way to localize the acceptor MOs on a specific part of a π-conjugated molecule is to substitute the backbone with acceptor groups. In Ref. [32], we have shown that a variation in the position of nitro substituents can be used to modify RR in a systematic way by exploiting the fact that the slopes $k$(MO) in figure 3 are intimately linked to the localization pattern of the respective MOs. This study concluded that the closer the nitro group is placed with respect to the electrodes, the higher is the RR.

In figure 6, we show the MO level shifts and bias-dependent transmission functions for another unimolecular rectifier substituted by nitro groups. The π-conjugated backbone in the center of the molecule is a so-called Tour wire, for which electron transport calculations have been carried out before, though with different chemical substitution patterns [23,33]. The HOMO of compound E is lying closer to the Fermi level than the LUMO in the NEGF-DFT calculations and dominates the asymmetry found with respect to the bias window for the two voltage polarities in figures 6b and c. Although we have added acceptor groups to an otherwise spatially symmetric molecular wire, it is thus the properties of the HOMO level that we have modulated to introduce a rectification in the I/V curve. The exact energetic alignment of the Fermi level with respect to the molecular frontier orbitals is governed by the charge transfer between the molecule and the electrodes at zero bias and depends on all aspects of the molecular structure and its attachment to the metal surface [40]. In general, it is always a simplification to pick out one of the eight parameters in figure 3 as the decisive factor for the rectification since it is the interplay of all these parameters that characterize the transport properties of an asymmetric molecule.

Regarding the level crossings related to electron jumps in figure 6a, it is the steepness of the slopes of the MOs involving the $NO_2$-groups (most notably the LUMO) which primarily defines Q(forward) and Q(reverse). The spatial distribution of the HOMO differs in figures 6a and b since it is localized on just one repeat unit of the Tour wire in the AM1 calculations, whereas it covers two benzene rings in the NEGF-DFT results. This is due to differences in the voltage range between the two figures since the orbital would have been delocalized also over two rings at the AM1 level if we had shown the orbital shape for zero or low bias. Note that the orbital shapes in both pictures refer to the voltages pinpointed by the arrows.

**4. Current-voltage curves, rectification ratios, and technological usefulness**

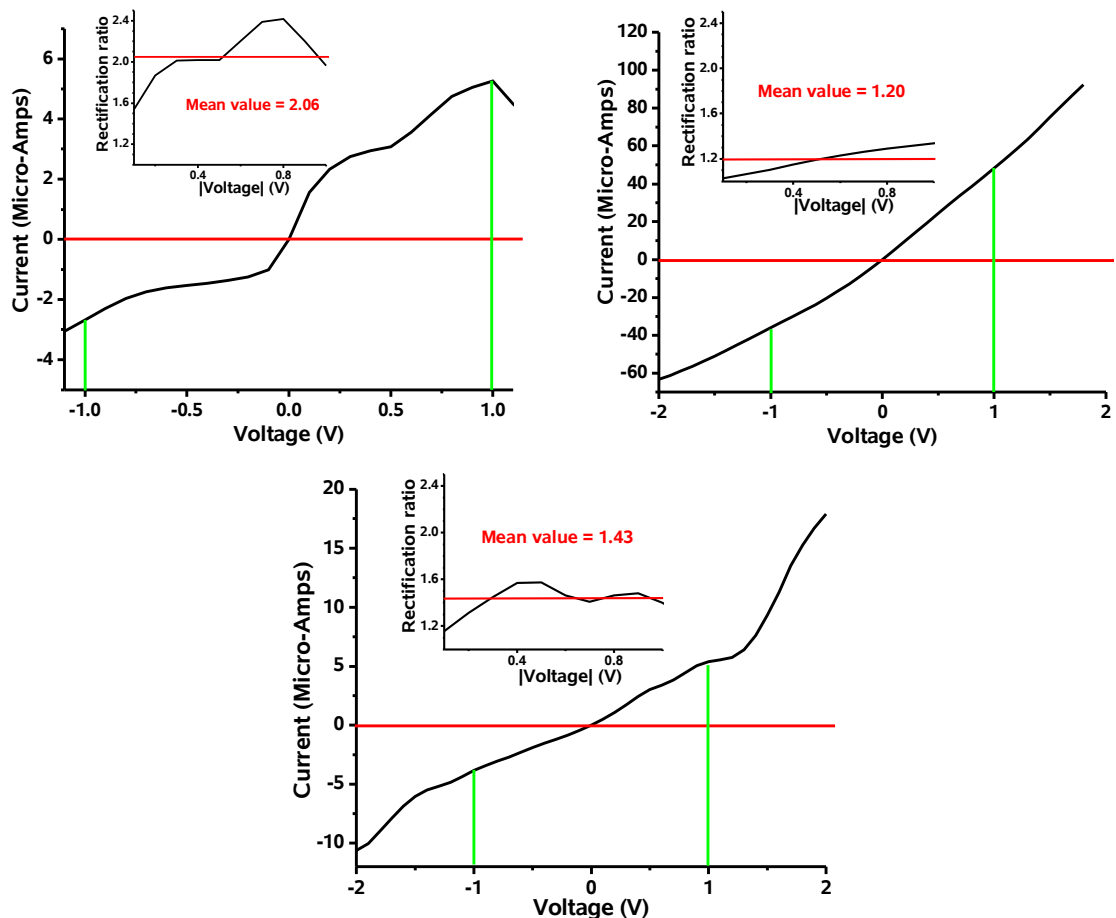

**Figure 7.** Current/voltage curves calculated with the NEGF-DFT method for molecules C (top left), D (top right) and E (bottom), respectively. The evolution of the rectification ratio (RR) as a function of the bias voltage and their average value are shown in the insets; for the three molecules, the averaging has been done from the data obtained from 0 to +/-1 V with steps of 0.1 V, as highlighted by the vertical lines in the I/V curves.

**Table 1.** Rectification ratios calculated for all unimolecular rectifiers investigated here from intra-molecular electron jumps as Q(reverse)/Q(forward) at the AM1 level and from I/V curves obtained with NEGF-DFT as an average of I(forward)/I(reverse) over the voltage range 0 to +/- 1V, with steps of 0.1 eV.

| Structure | AM1  | NEGF-DFT |
|-----------|------|----------|
| A         | 2.25 | -        |
| B         | 2.20 | -        |
| C         | 1.81 | 2.06     |
| D         | 1.80 | 1.20     |
| E         | 1.48 | 1.43     |

A key quantity for evaluating whether a single molecule diode can be used for practical applications as an electronic device is the rectification ratio, which distinguishes the ON state from the OFF state for the current flow. For most purposes, a rather high value (typically larger than 1000) is required for RR. We stress that RR is a figure of merit rather than a well-defined physical property of a nano-junction; its exact definition (which varies in the literature) can also depend on the characteristics of a particular device setup. RR is usually derived from experimental I/V curves such as those calculated with NEGF-DFT for structures C, D, and E, see figure 7. In Table 1, we compare the values of RR extracted from the I/V curves to those estimated from the ratio of the threshold sparkle charges with AM1. For the five molecules, RR is very low and far away from technological usefulness.

The two definitions employed for RR in Table 1 differ significantly in the sense that one is an average over a ratio of currents (through electrode/molecule/electrode junctions) and the other a ratio of critical voltages for intra-molecular electron jumps. It is therefore striking to see that the values derived from NEGF-DFT and AM1 correlate rather well, with the exception of the zwitterion (molecule D) for which no level crossings could be found for the intra-molecular electron jump in the forward bias direction with AM1 (the jump in the reverse direction occurs at such a high voltage that it is irrelevant for a comparison with the low-bias NEGF-DFT study). This overall agreement of the two methods is even more remarkable when observing in figure 7 that the three junctions exhibit very different shape in the I/V curves; the latter is step-like for molecule C, perfectly continuous for D, and shows an intermediate behaviour for E. The three curves also differ significantly in the order of magnitude of the current for a given bias. Both features are related to the strength of the chemical coupling of the involved MOs to the surface states of the electrodes and can be rationalized from figures 4-6. For molecules C and D, the transmission functions are driven by the vicinity of the HOMO level to the bias window for all displayed values of the voltage; what makes them different is the width of the HOMO peak in the transmission spectrum. For molecule C, the HOMO is localized on the donor side of the molecule and is weakly coupled to at least one of the electrodes, thus leading to a narrow transmission peak. The current has a sharp onset when this peak slips into the bias window and its magnitude is related to the integral of the transmission over the bias window. For molecule D, the coupling of the frontier orbitals to the electrodes is large, and so is the width of the transmission peaks, whose tails fill up the bias window to result into a large current. In this case, the increase of the current with the bias can be mainly attributed to the enlargement of the window of integration, thus rationalizing the low value calculated for RR. Molecule E is a special case where the strongly coupled HOMO governs the transmission characteristics but is energetically further apart from the Fermi level compared to molecules C and D. If the forward bias goes beyond +1.2V, the main peak at -1 eV in the transmission function fully enters the integration window and induces an abrupt increase in the current.

There are recent experimental reports for RR values of 50-150 obtained for self-assembled monolayers made of sterically hindered dyes [19]. Since the molecules investigated by Ashwell *et al.* in Ref. 19 are asymmetric in their capacitive coupling to a gold surface and a platinum tip and contain in addition a rather long alkyl chain on the surface side, it is not straightforward to assess theoretically the amount of rectification arising from the intramolecular charge separation within the dye. On the basis of tight-binding calculations, a value of RR ~500 has been proposed by using alkyl spacers of different lengths [35]; note however that this value has been downscaled to ~35 when cross-checked with DFT calculations by the same group [41]. This small ratio is nevertheless rather high compared to the results obtained in our present study. The explanation for that is two-fold: i) the asymmetry was introduced in Ref. 35 by alkyl spacers of different lengths and has not a truly unimolecular origin; ii) the alkyl spacers decouple the central moiety from the electrodes, resulting in sharp steps in the I/V curves. The authors exploit this fact by adopting a definition for RR which is in the spirit of a Zener diode, namely the current ratio is not averaged over the whole bias range but is estimated only at a carefully chosen operating voltage.

## 5. Concluding remarks

We have presented a theoretical survey of the rectification properties of three different classes of single molecule rectifiers and introduced a general analysis scheme based on the bias dependence of the energetic positions of the key molecular orbitals. In most cases, a good estimate of the rectification ratio can be obtained by studying the threshold voltages required for intra-molecular one-electron jumps by using the cost-effective parameterized AM1 technique ; however, the more sophisticated NEGF-DFT method has to be employed to gain a deeper understanding of the physics of the electron transport through nano-junctions and the role of the coupling to the electrodes. Our general analysis scheme differs when compared to earlier theoretical frameworks set up to characterize unimolecular rectifiers [1],[22] since we identify here a set of eight parameters to be taken into account, namely the energies at zero bias of the frontier orbitals on the donor and acceptor sides and the slopes of their evolution under the influence of an external electric field. Depending on the system under investigation, any of these parameters can be crucial for the asymmetry found in the I/V curves.

For all studied molecules, RR was found to be far too low for any practical application and we suspect this might be the case for many donor-acceptor systems ("push-pull compounds"). AM1 calculations appear to be a good method for screening a large number of asymmetric molecules for their rectification ratios. It is, however, a formidable challenge to conduct such a search systematically, since RR depends on the energetic positions and the field evolution of four molecular orbitals. These eight variables cannot be independently tuned by varying chemical structures and we have evidences that what can be gained on RR by optimizing one of these variables is in most cases lost by changing simultaneously the values of the other parameters. A recent theoretical study based on a double barrier tunneling model also came to the conclusion that for coherent transport an upper limit of RR~22 has to be expected [42]. Therefore, it would be of prime interest for this field of research to go beyond purely electronic processes in the strong coupling regime of coherent tunneling and focus on proposals for diodes based on a more complex physical origin, such as Coulomb blockade behaviour [43], Debye-screening in electrochemical systems [44], multi-phonon suppression [45] or intermolecular non-adiabatic electron transfer processes [46].

## Acknowledgements

The authors acknowledge very stimulating discussions with Dr. D. De Leeuw. The work in Mons has been supported by the European Integrated Project NAIMO (NMP4-CT-2004-500355), the Interuniversity Attraction Pole IAP 6/27 Program of the Belgian Federal Government « Functional supramolecular systems (FS2) », and the Belgian National Fund for Scientific Research (FNRS). Access to the software of Atomistix Inc. has been provided from a collaboration within the EC STREP project MODECOM (NMP-CT-2006-016434); we are especially indebted to Dr. J. A.Torres for his advices for applying this method. J.C. is a Research Associate of FNRS.

A general theoretical view on unimolecular rectifiers     - 17 -[44] Kornyshev A A, Kuznetsov A M and Ulstrup J 2006 *Proc. Nat. Ac. Sci.* **103** 6799
[45] Petrov E G, Zelinsky Y R, May V and Hänggi P 2007 *J. Chem. Phys.* **127** 084709
[46] Broo A and Zerner M C 1995 *Chem. Phys.* **196** 423